\documentclass[12pt]{iopart}
\usepackage{graphicx}
\usepackage{cite}

\def\la{\langle}
\def\ra{\rangle}

\def\l{\left}
\def\r{\right}

\def\dgbs{\Delta\Gamma_{B_s}}
\def\dgog{{\dgbs\over\Gamma_{B_s}}}
\def\fbs{f_{B_s}}
\def\MBs{M_{B_s}}

\def\beq{\begin{equation}}
\def\eeq{\end{equation}}
\def\bea{\begin{eqnarray}}
\def\eea{\end{eqnarray}}

\def\gev{\mbox{ GeV}}
\def\mev{\mbox{ MeV}}
\def\msbar{\overline{\mbox{MS}}}

\newcommand{\op}{{\mathcal{O}}}

\begin{document}

\begin{flushright}
SHEP 00/14\\
\end{flushright}

\title[Mixing and lifetimes from lattice QCD]
{$B^{0}_{(s)}{-}\bar{B}^{0}_{(s)}$ Mixing and
$b$ Hadron Lifetimes\\from Lattice QCD\footnote{Based on talks given by
the authors in the {\it UK Phenomenology Workshop on Heavy Flavours
and CP Violation}, 17th - 22nd September 2000, Durham, England.}}

\vspace{0.7cm}

\author{Jonathan Flynn and C.-J. David Lin}

\address{Department of Physics and Astronomy, the University 
of Southampton, Southampton SO17 1BJ, UK}

\begin{abstract}
We discuss neutral $B_{s,d}$ meson mixing and $b$-hadron lifetimes
from the perspective of recent lattice calculations. In particular, we
consider matrix elements which can be combined with measured $\Delta
M_{s,d}$ to constrain $|V_{td}|$, the lifetime ratios
$\tau(\Lambda_b)/\tau(B^0_d)$ and $\tau(B^-)/\tau(B^0_d)$ and the
lifetime difference, $\dgbs/\Gamma_{B_s}$, in the neutral $B_s$ meson
system.
\end{abstract}



\maketitle

\section{Introduction}

In this workshop, 
the lifetimes and mixings working group identified a `wish list' of
hadronic quantities whose accurate theoretical determination is
crucial for understanding current and future $b$-physics experimental
results.  We discuss several of these quantities from the perspective
of recent lattice calculations:
\begin{itemize}
\item
matrix elements which can be combined with measured $\Delta M_{s,d}$
to constrain $|V_{td}|$
\item
the lifetime ratios $\tau(\Lambda_b)/\tau(B^0_d)$ and
$\tau(B^-)/\tau(B^0_d)$
\item
the lifetime difference, $\dgbs/\Gamma_{B_s}$, in the neutral $B_s$
meson system.
\end{itemize}
For a full survey of recent $b$-physics results from the lattice see
the reviews in
\cite{BernardLatt2000,HashimotoLatt99,Draper,FlynnSachrajda,
WittigVarenna,cthd}.

\section{$\Delta M_{d}$ and $\Delta M_{s}/\Delta M_{d}$ }
\label{sec:dmds}

The mass difference of the $B_d$--$\bar B_d$ system, $\Delta M_d$, 
constrains the poorly known CKM matrix element $|V_{td}|$.
$\Delta M_d$ has been experimentally measured to good accuracy.
On the theory side, the main uncertainty comes from the long-distance
strong-interaction effects in the matrix element
\beq
\label{eq:me_def}
{\mathcal{M}}_{B_{d}}(\mu) = 
\la \bar B_d|Q_{L_{d}}(\mu)|B_d\ra ,
\eeq
which appears in the Standard Model prediction for $\Delta M_{d}$ to
leading order in an expansion in $1/M_W$ 
\cite{buras:1990fn, buchalla:1996vs}.  In
Eq.~\eref{eq:me_def}, $Q_{L_{d}}$ is the four-quark operator $[\bar
b\gamma^\mu(1-\gamma^5)d] [\bar b\gamma_\mu(1-\gamma^5)d]$ and $\mu$
is the renormalisation scale.

An alternative approach, in which many theoretical
uncertainties cancel, is to consider the ratio, $\Delta M_s/\Delta
M_d$, where $\Delta M_s$ is the mass difference in the neutral
$B_s-\bar B_s$ system. In the Standard Model, one
has
\beq
\label{eq:deltamsovermd}
\frac{\Delta M_s}{\Delta M_d} = 
\l|\frac{V_{ts}}{V_{td}}\r|^2\l(\frac{M_{B_d}}{M_{B_s}}\r)
\l|\frac{\la \bar B_s|Q_{L_{s}}|B_s\ra}
{\la\bar B_d|Q_{L_{d}}|B_d\ra}\r|
\equiv
\l|\frac{V_{ts}}{V_{td}}\r|^2\l(\frac{M_{B_s}}{M_{B_d}}\r)\xi^2
\ ,
\eeq
where $Q_{L_{s}}$ is the same operator as
$Q_{L_{d}}$ with $d$ replaced by $s$ and where 
the renormalisation-scale dependence of these operators
cancels in the ratio. Because the unitarity of the CKM matrix
implies $|V_{ts}|{\simeq} |V_{cb}|$ and because $|V_{cb}|$ can be
accurately obtained from semileptonic $B$ to charm decays, a
measurement of $\Delta M_s/\Delta M_d$ determines 
$|V_{td}|$. This is experimentally very challenging because of the
rapid oscillations in the $B^{0}_{s}{-}\bar{B}^{0}_{s}$ system.  Nevertheless,
the experimental lower bounds on $\Delta M_s/\Delta M_d$ already yield
interesting constraints on the $b{-}d$ unitarity triangle 
\cite{StocchiDurham00}.

The matrix elements in Eqs.~\eref{eq:me_def} and
\eref{eq:deltamsovermd} are traditionally parameterised by
\beq
\label{eq:bparamdef}
{\mathcal{M}}_{B_q}(\mu) =
\la\bar B_q|Q_{L_{q}}(\mu)|B_q\ra
= \frac{8}{3} M_{B_q}^2 f_{B_q}^2 B_{B_q}(\mu)
\ ,
\eeq
with $q=d$ or $s$, where the $B$-parameter, $B_{B_q}$, measures
deviations from vacuum saturation, corresponding to $B_{B_q}=1$, and
$f_{B_q}$ is the leptonic decay constant.  One also usually introduces 
a renormalisation-group invariant and scheme-independent parameter
$\hat{B}_{B_{q}}$, which to NLO in QCD is given by
\beq
 \hat{B}^{\mathrm{nlo}}_{B_{q}} = C_{B}(\mu) B_{B_{q}}(\mu) ,
\eeq
where $C_{B}$ is the two-loop Wilson coefficient calculated in the
same scheme as the matrix element \cite{buras:1990fn,Ciuchini:1995cd}.

Because $M_{B_s}$ and $M_{B_d}$ are measured experimentally, one 
needs to calculate $\xi$ and 
$f_{B_{d}}\sqrt{\hat{B}^{\mathrm{nlo}}_{B_{d}}}$
non-perturbatively to determine $|V_{td}|$ from experimental
results for $\Delta M_{s}/\Delta M_{d}$ and 
$\Delta M_{d}$.

There are three recent quenched lattice calculations of $\xi$
\footnote{The $SU(3)$ breaking ratio 
${\mathcal{M}}_{B_s}/{\mathcal{M}}_{B_d}$
has also been studied in \cite{BerBlumSoni,LellouchLinBBAll} by using
a different method, in which one does not need to calculate
$\xi$.  However, results from this method have large uncertainties.}.
The APE~\cite{Becirevic:2000nv} and UKQCD~\cite{LellouchLinBBAll} (with
preliminary results reported previously in \cite{LellouchLinBBPreli})
Collaborations use relativistic formulations of quarks and obtain
$\xi$ at the physical $B$ meson mass by extrapolating from 
heavy-meson masses around that of the $D$.  Gim\'{e}nez and Martinelli
(denoted as GM in the following) \cite{GimMart} calculate this
quantity in the static limit where the $b$ quark mass is taken to
infinity.  The results from these three groups are:
\beq
\label{eq:xiResults}
 \xi = \l\{ \begin{array}{l} 1.16(7)\mbox{ }\mbox{ }\mbox{ }
  \mbox{ }\mbox{ }\mathrm{(APE)}\\ 
        1.16(2)^{+2}_{-3}\mbox{ }\mbox{ }\mathrm{(UKQCD)}\\ 
        1.17(3)\mbox{ }\mbox{ }\mbox{ }
        \mbox{ }\mbox{ }\mathrm{(GM)} , \\ \end{array} \r . 
\eeq
where the first error is statistical and the second is systematic.
In order to avoid the situation where discretisation errors, which 
can be significant in lattice calculations involving propagating
heavy quarks, 
are out of control, both APE and UKQCD perform numerical 
simulations
with several meson masses straddling the $D$ and then extrapolate
to the $B$ meson mass using HQET-inspired relations.
APE use a fully $\op(a)$ ($a$ is the lattice spacing) improved 
fermion action \cite{NPcSW}
in which the leading discretisation error is $\op(a^2)$.
UKQCD use a mean-field improved fermion action \cite{GLeM}, 
and the leading discretisation error here is {\it formally} 
$\op(a\alpha_{s})$, although it might be numerically smaller.
However, it should be noted that
both groups have $\op(a\alpha_{s})$
discretisation errors in $\xi$, because the 
four-quark operators $Q_{L_{q}}$ are not fully $\op(a)$ 
improved in these calculations.  These $\op(a\alpha_{s})$ errors are 
absent in the decay constants, and hence $f_{B_{s}}/f_{B_{d}}$,
as calculated by APE, because they use the improved 
lattice axial current in the calculation.

In Eq.~\eref{eq:xiResults}, UKQCD's result for $\xi$ has a much 
smaller statistical error than that obtained by APE, although
these two groups have very similar statistics in the Monte-Carlo
simulations.  This is because UKQCD calculate $\xi$ by performing
heavy-quark-mass extrapolations in the ratios 
$f_{B_{s}}\sqrt{M_{B_{s}}}/f_{B_{d}}\sqrt{M_{B_{d}}}$ 
and $B_{B_{s}}/B_{B_{d}}$, in which the $1/m_{Q}$ corrections
cancel significantly, especially for the decay constants
\cite{LellouchLinBBAll,LellouchLinBBPreli}.  This procedure
determines $\xi$ to a better statistical accuracy than that
calculated by extrapolating $f_{B_{s}}\sqrt{M_{B_{s}}}$, 
$f_{B_{d}}\sqrt{M_{B_{d}}}$, $B_{B_{s}}$, and $B_{B_{d}}$
individually.

The above three studies suggest that $\xi$ has small systematic
uncertainties arising from 
discretisation effects and heavy-quark-mass extrapolation\footnote{As
mentioned above, the
authors of \cite{LellouchLinBBAll} look at the $1/m_{Q}$ corrections
for $(f_{B_{s}}\sqrt{M_{B_{s}}})/(f_{B_{d}}\sqrt{M_{B_{d}}})$ and
$B_{B_{s}}/B_{B_{d}}$ and find that they are very small for both
quantities.}, within the quenched approximation.  
The UKQCD result includes a systematic error\footnote{This systematic
error looks small on $\xi$.  However, it should be noted that 
it is the error in $\xi$'s deviation from unity.} not
estimated by APE and GM.  This is due to the uncertainty in the
determination of the lattice spacing (see below for details), which
introduces a variation of the strange-quark mass.

Quenched Chiral Perturbation Theory (q$\chi$PT) predicts that 
quenching errors in $B_{B_{s}}/B_{B_{d}}$ are small if the 
couplings in the theory are constrained by large-$N_{c}$
arguments \cite{Sharpe:1996qp}.  Recent numerical studies with
two flavours of dynamical quarks show little variation in
$f_{B_{s}}/f_{B_{d}}$ compared to its quenched value
\cite{Bernard:1999nv,Collins:1999ff,Khan:2000eg,Bernard:2000unq}.  
Therefore,
quenching effects should not be significant in $\xi$.  For more
detailed discussions on this issue, please refer to 
\cite{mcneile}.

APE~\cite{Becirevic:2000nv} and UKQCD~\cite{LellouchLinBBAll} also
calculate $f_{B_{d}}\sqrt{\hat{B}^{\mathrm{nlo}}_{B_{d}}}$ in quenched
approximation.  The results are:
\beq
\label{eq:fBBBResults}
 f_{B_{d}}\sqrt{\hat{B}^{\mathrm{nlo}}_{B_{d}}} = \l\{ \begin{array}{l}
 206(28)(7)\mbox{ }\mev\mbox{ }\mbox{ }\mbox{ }\mathrm{(APE)}\\
 211(21)(28)\mev\mbox{ }\mbox{ }\mbox{ }\mathrm{(UKQCD)} ,
 \\ \end{array} \r . 
\eeq
where the first error is statistical and the second is systematic.

The systematic error in APE's result for
$f_{B_{d}}\sqrt{\hat{B}^{\mathrm{nlo}}_{B_{d}}}$ reflects the typical
size of the $\op(a^2)$ discretisation effects in $f_{B_{q}}$, a very
small ($\sim 2\%$) uncertainty in the inverse lattice spacing and
the uncertainty in matching the lattice-regularised four-quark
operators onto the NDR-$\msbar$ scheme, with these three errors taken
in quadrature.

In addition to the $\op(a^2)$ and $\op(a\alpha_{s})$ discretisation
errors in $f_{B_d}$, UKQCD investigate the uncertainties in the procedure of
operator matching, heavy-quark-mass extrapolations 
and a $\pm 7\%$ uncertainty in the inverse
lattice spacing.  The systematic error in UKQCD's result is obtained
by taking these uncertainties in quadrature.

The overall systematic error on
$f_{B_{d}}\sqrt{\hat{B}^{\mathrm{nlo}}_{B_{d}}}$ obtained by UKQCD is
much larger than that obtained by APE (see Eq.~\eref{eq:fBBBResults}),
mainly because of the estimate of
the error in the determination of the inverse lattice spacing,
$a^{-1}$.  APE use $a^{-1}$ set by the method described in
\cite{Allton:1997yv} for their central
value, and the ones set by $M_{\rho}$ and $M_{\phi}$ to get the systematic
uncertainty.  This method results in a $2\%$ systematic error.  UKQCD
determine $a^{-1}$ in conjunction with the strange-quark mass 
by $f_{K}$ and $M_{K}$, and then vary $a^{-1}$ by $\pm 7\%$, a
range which covers the typical variations of $a^{-1}$ set by gluonic
and light-hadron spectral quantities with the same action used in
their calculation \cite{Bowler:1999ae}.  This procedure introduces a
$9\%$ effect on their result for
$f_{B_{d}}\sqrt{\hat{B}^{\mathrm{nlo}}_{B_{d}}}$.  In addition, the
systematic uncertainty from the heavy-quark-mass extrapolations 
($\sim3\%$)
which is not estimated by APE, and the $\op(a\alpha_s)$ discretisation
errors in $f_{B_d}$ ($\sim5\%$) which are not present in the APE
calculation, further enlarge the gap.

\begin{figure}
\begin{center}
\includegraphics[width=0.7\textwidth]{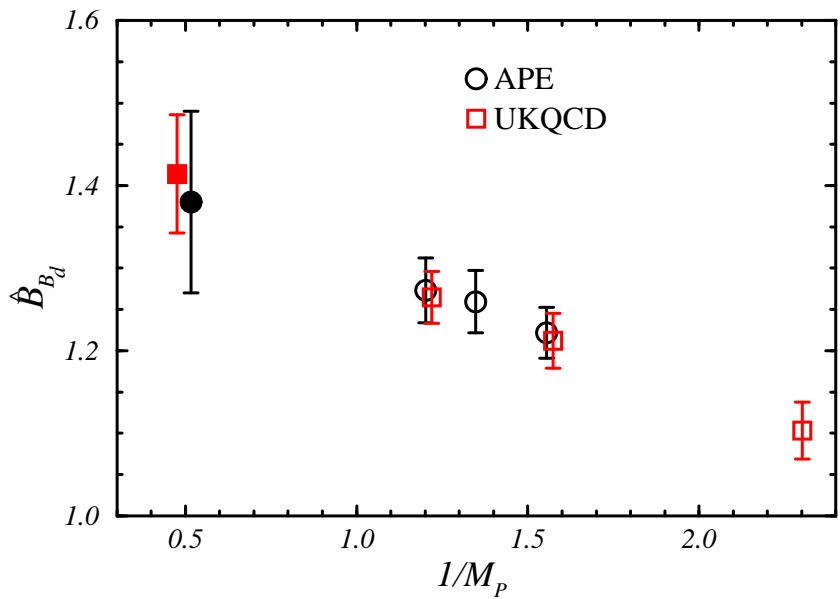}
\end{center}
\caption{\label{fig:APEvsUKQCD} {Comparison between APE and UKQCD's
results of renormalisation group invariant $B$ parameters at different
meson masses (from \cite{Becirevic:2000nv}).  The inverse lattice
spacing in this plot is approximately $2.7\gev$.  $M_{P}$ is the
heavy-meson mass in lattice units ({\it i.e.,} one should multiply it
by 2.7 to obtain the mass in GeV in this case).  The open symbols are
the results from numerical simulations, and the closed symbols are
those extrapolated to the $B_{d}$ meson mass.}}
\end{figure}

Both APE and UKQCD do not attempt to quantify the discretisation
errors in the $B$ parameters, as this requires one to consider the
mixing with dimension-seven four-fermion operators and is therefore
very involved.  Nevertheless, $B_{B_{q}}$ are ratios between closely
related matrix elements, hence one expects that these errors might
cancel significantly.  As mentioned above, the fermion actions used by
APE and UKQCD have different discretisation errors.  However, as
displayed in Table \ref{tab:Bparam} and Figure \ref{fig:APEvsUKQCD},
their results of the $B$ parameters show very good agreement.
Furthermore, UKQCD perform the calculations at two lattice spacings
and observe that the change in $B_{B_{q}}$ is insignificant.  Finally,
both groups obtain $B_{B_{q}}$ compatible with those in a slightly 
earlier work by Bernard, Blum and Soni (BBS) \cite{BerBlumSoni}, who use a
less improved fermion action but extrapolate their results to the
continuum limit (See Table \ref{tab:Bparam}).  Recently, these $B$
parameters have also been calculated using NRQCD to describe the heavy
quarks \cite{Hiroshima,yamada:2000ym}, with the
heavy-quark masses in numerical simulations around that of the $b$.
The central values of these results are 2 standard-deviations lower
than those obtained by APE and UKQCD.  However, as Table
\ref{tab:Bparam} shows, when the systematic errors are taken into
account, all these results are compatible.
\begin{table}
\begin{center}
\caption{\label{tab:Bparam} {Comparison of the latest lattice results
for $B_{B_{q}}$ parameters
renormalised in NDR-$\msbar$ scheme at $m_{b}$.   
The first error on each
result is statistical, while the second one is systematic.  For a more
complete list of these results, including the earlier calculations,
please refer to 
\cite{BernardLatt2000,HashimotoLatt99,Draper,FlynnSachrajda,
WittigVarenna,cthd}.}}
\begin{tabular}{ccc}
\hline
\hline
 & $B_{B_{d}}(m_{b})$ & $B_{B_{s}}(m_{b})$\\
\hline
APE ($a^{-1}\approx 2.7\gev$) \cite{Becirevic:2000nv} & $0.93(8)^{+0}_{-1}$ 
& $0.92(3)^{+0}_{-1}$\\
UKQCD ($a^{-1}\approx 2.7\gev$) \cite{LellouchLinBBAll} & $0.92(4)^{+3}_{-0}$ 
& $0.91(2)^{+3}_{-0}$\\
UKQCD ($a^{-1}\approx 2.0\gev$) \cite{LellouchLinBBAll} & $0.90(4)^{+3}_{-0}$
& $0.92(2)^{+3}_{-0}$\\
BBS ($a^{-1}\rightarrow\infty$) \cite{BerBlumSoni} & 0.95(12) & 0.95(12) \\
NRQCD ($a^{-1}\approx 2.3\gev$) 
\cite{Hiroshima, yamada:2000ym} & 0.84(2)(8) & 0.87(1)(9)\\
\hline
\hline
\end{tabular}
\end{center}
\end{table}

UKQCD also normalise $f_{B_{d}}\sqrt{\hat{B}^{\mathrm{nlo}}_{B_{d}}}$
by $f_{D_{s}}$ because some systematic errors cancel, and there have
been preliminary experimental results of $f_{D_{s}}$ \cite{fDsEXP1,fDsEXP2}.
They obtain \cite{LellouchLinBBAll}:
\beq
 \frac{f_{B_{d}}}{f_{D_{s}}}\sqrt{\hat{B}^{\mathrm{nlo}}_{B_{d}}} =
 0.89(7)^{+6}_{-6} ,
\eeq
where the first error is statistical and the second one is systematic.

While quenching errors in $B_{B_{q}}$ are predicted to be within a few
percent by q$\chi$PT with reasonable ranges of the couplings in the
theory \cite{Sharpe:1996qp}, they can be significant for the decay
constants.  Recent numerical studies with two flavours of dynamical
quarks
\cite{Bernard:1999nv,Collins:1999ff,Khan:2000eg,Bernard:2000unq} show
visible increases on the decay constants.  Again, please refer to
\cite{mcneile} for more detailed discussions on quenching errors.

Finally, there are on-going analyses 
\cite{BernardLatt2000,grb,UKQCDandAPEBB} in
which results from the static-limit and relativistic-heavy-quark
calculations are combined.  In such analyses, 
$f_{B_{d}}\sqrt{\hat{B}^{\mathrm{nlo}}_{B_{d}}}$ and $B_{B_{q}}$
are obtained by interpolations between the charm-mass region and
the limit of infinite heavy-quark mass.

\section{Spectator Effects on $b$-Hadron Lifetimes}

The combination of operator product and heavy quark expansions
predicts that the ratio of lifetimes of two hadrons, $H_1$ and $H_2$,
each containing a single $b$-quark, is given by \cite{ns96},
\begin{equation}
{\tau(H_1)\over \tau(H_2)} = 1 +
 {\mu_\pi^2(H_1) - \mu_\pi^2(H_2)\over 2m_b^2} + 
 c_{\mathrm{G}}
    {\mu_{\mathrm{G}}^2(H_1) - \mu_{\mathrm{G}}^2(H_2)\over m_b^2} +
 O(1/m_b^3).
\end{equation}
Here the leading $1$ on the right hand side arises from a universal
first term describing free $b$-quark decay. There is no term of order
$1/m_b$ in the expansion. At order $1/m_b^2$, the $\mu_\pi^2$ and
$\mu_{\mathrm{G}}^2$ terms are matrix elements in the heavy quark
effective theory of the kinetic energy and chromomagnetic moment
operators respectively between the corresponding hadron states. They
can be fixed from hadron mass formulas and mass splittings. The
coefficient $c_{\mathrm{G}}$ is calculated in
\cite{Bigi:1992su,Bigi:1993fe,Bigi:1993proc,Falk:1994gw}. 
Therefore, for the cases
of the ratios of $B^-$ and $B^0_d$ lifetimes or $\Lambda_b$ and
$B^0_d$ lifetimes, we have calculated values to compare to
experimental results:
\begin{equation}
{\tau(B^-)\over \tau(B^0_d)} = \cases{1 + O(1/m_b^3)&\\
                              1.066(20)\mbox{\ \cite{lephfs}}&\\}
\qquad
{\tau(\Lambda_b)\over \tau(B^0_d)} = \cases{0.98 + O(1/m_b^3)&\\
                              0.794(53)\mbox{\ \cite{lephfs}}&\\}
\end{equation}
For the $B$-meson ratio, there is no difficulty. However, for
the $\Lambda_b$ to $B$ ratio, the $O(1/m_b^3)$ terms apparently must
account for almost all of the deviation from one.

\begin{figure}
\begin{center}
\def\point#1 #2 #3{\put(#1,#2){\makebox(0,0){\small$#3$}}}
\unitlength0.0018\textwidth
\begin{picture}(330,154)
\put(0,0){\includegraphics[width=330\unitlength]{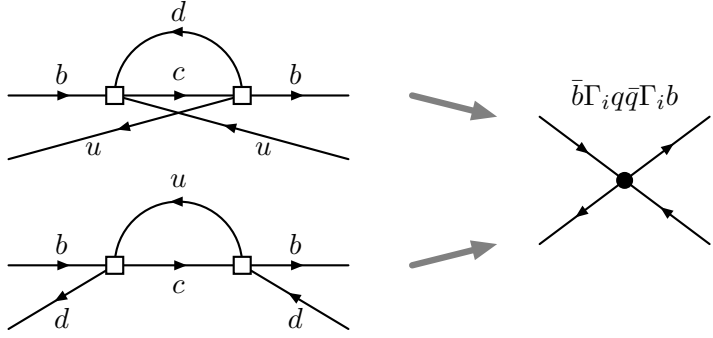}}
\point 25 5 d
\point 135 5 d
\point 25 40 b
\point 135 40 b
\point 80 20 c
\point 80 70 u
\point 40 85 u
\point 120 85 u
\point 25 120 b
\point 135 120 b
\point 80 120 c
\point 80 150 d
\point 290 110 {\bar b \Gamma_i q \bar q \Gamma_i b}
\end{picture}
\end{center}
\caption[]{Spectator contributions: diagrams like those on the left,
where the box denotes a $|\Delta B| =1$ transition, produce 4-quark
operators in the operator product expansion. $\Gamma_i$ denotes some
combination of Dirac and colour matrices. See~\cite{ns96}.}
\label{fig:specteffects}
\end{figure}

Some work has addressed the question of whether ``spectator effects''
in the $O(1/m_b^3)$ terms can be responsible for such large
deviations. In the operator product expansion, the effects of the
diagrams in figure~\ref{fig:specteffects} lead to the appearance of
$\Delta B =0$ four-quark operators of the form $\bar b \Gamma_i q \bar
q \Gamma_i b$. These diagrams are the first time the effects of
spectator quarks enter explicitly. Moreover, the diagrams are
one-loop, compared to the two-loop diagrams which generate
$O(1/m_b^2)$ terms in the expansion. There is thus a hope that these
spectator effects could be large. The four quark operators are
classified as
\begin{eqnarray}
O^q_1 = \bar b_L \gamma_\mu q_L \bar q_L \gamma^\mu b_L \qquad &
 T^q_1 = \bar b_L \gamma_\mu T^a q_L \bar q_L \gamma^\mu T^a b_L \\
O^q_1 = \bar b_R q_L \bar q_L b_R &
 T^q_2 = \bar b_R T^a q_L \bar q_L T^a b_R
\end{eqnarray}
where $q=u,d,s$. The spectator contribution to the decay rate of a
hadron $H$ depends on $\langle H | O_{\mathrm{spec}}| H \rangle$,
where
\begin{equation}
O_{\mathrm{spec}} = F_{u1}(z) O^u_1 + G_{u1}(z) T^u_1 + 
 \sum_{i=1,2; q=d,s}\left(F_{iq}(z)O^q_i + G_{iq}(z)T^q_i \right).
\end{equation}
Here $z=m_c^2/m_b^2$ and the $F$'s and $G$'s are coefficient functions
known at leading order~\cite{ns96}. 

Given the coefficient functions, the remaining ingredient is the
evaluation of the four quark matrix elements.  Lattice QCD simulations
of these~\cite{mdpcts98,mdpctslat98,mdpctscmi} are reported here (for
sum rule calculations see~\cite{defazioa,defaziob}). To date, these
calculations are still exploratory for the $\Lambda_b$ meson, but the
results are encouraging and the calculations deserve repeating.

It is convenient to parameterise the matrix elements. For mesons we use
factors $B_i$ and $\epsilon_i$,
\begin{equation}
{1\over2 m_B} \langle B | O^q_i | B \rangle = {f_B^2 m_B\over 8} B_i,
\qquad
{1\over2 m_B} \langle B | T^q_i | B \rangle = {f_B^2 m_B\over 8} \epsilon_i,
\end{equation}
for $i=1,2$. From vacuum saturation or large $N_c$ arguments the
expectation is that $B_i \simeq 1$ and the $\epsilon_i$ are small. For
the $\Lambda_b$ baryon, heavy quark symmetry implies that only two
matrix elements need to be considered (up to $1/m_b$
corrections)~\cite{ns96}. The corresponding parameters, $L_1$ and
$L_2$ are defined by\footnote{The conversion to the parameters used
in~\cite{ns96} is $\tilde B = -6L_1$ and $r = -2 L_2/L_1 - 1/3$.},
\begin{equation}
{1\over2 m_B} \langle \Lambda_b | O^q_1 | \Lambda_b \rangle =
  {f_B^2 m_B\over 8} L_1,
\qquad
{1\over2 m_B} \langle \Lambda_b | T^q_1 | \lambda_b \rangle =
 {f_B^2 m_B\over 8} L_2.
\end{equation}
In terms of these parameters the lifetime ratios can be expressed as,
\begin{eqnarray}
{\tau(B^-)\over \tau(B^0_d)} = 1 + a_1 \epsilon_1 + a_2\epsilon_2 +
  a_3 B_3 + a_4 B_4,\\
{\tau(\Lambda_b)\over \tau(B^0_d)} = 0.98 + b_1\epsilon_1 +
  b_2\epsilon_2 + b_3 L_1 + b_4 L_4,
\end{eqnarray}
where tiny $B_{1,2}$ terms are neglected in the second equation. The
coefficients $a_i$ and $b_i$ are known perturbatively,
\begin{equation}
\begin{array}{rcrrcr}
a_1 &=& -0.697 & \qquad b_1 &=& -0.175 \\
a_2 &=&  0.195 & \qquad b_2 &=&  0.195 \\
a_3 &=&  0.020 & \qquad b_3 &=&  0.030 \\
a_4 &=&  0.004 & \qquad b_4 &=& -0.252
\end{array}
\end{equation}
These values are quoted in the $\overline{\mathrm{MS}}$ scheme at a
scale $\mu=m_b$ and are to be combined with matrix elements evaluated
in the same scheme.

For the $B$ meson, the lattice calculations~\cite{mdpcts98} have been
done in the quenched approximation with a static $b$ quark at an
inverse lattice spacing of $a^{-1} = 2.9\gev$ (corresponding to an
input lattice coupling parameter $\beta=6.2$). The results have been
extrapolated from the light quark masses actually simulated (for
technical reasons, one cannot simulate with realistically light quark
masses) to the chiral limit. For the $\Lambda_b$ meson, the matrix
element involves a more complicated set of lattice quark propagator
contractions, as shown in figure~\ref{fig:contractions}, and the
calculation is still exploratory~\cite{mdpctslat98,mdpctscmi}. It is
also done with a static $b$ quark, but on a coarser lattice
($a^{-1}=1.1\gev$ or $\beta=5.7$) and using a stochastic technique to
calculate the light quark propagators~\cite{cmstoch}. No chiral limit
is taken in this case, the results are quoted for pion masses given by
$am_\pi = 0.52(3)$ (case A) and $am_\pi = 0.74(4)$ (case B). The
results are:
\begin{figure}
\begin{center}
\def\point#1 #2 #3{\put(#1,#2){\makebox(0,0){%
\vrule depth1.2ex width0pt height0pt\small$#3$}}}
\unitlength0.0018\textwidth
\begin{picture}(385,102)
\put(-20,0){\includegraphics[width=380\unitlength]{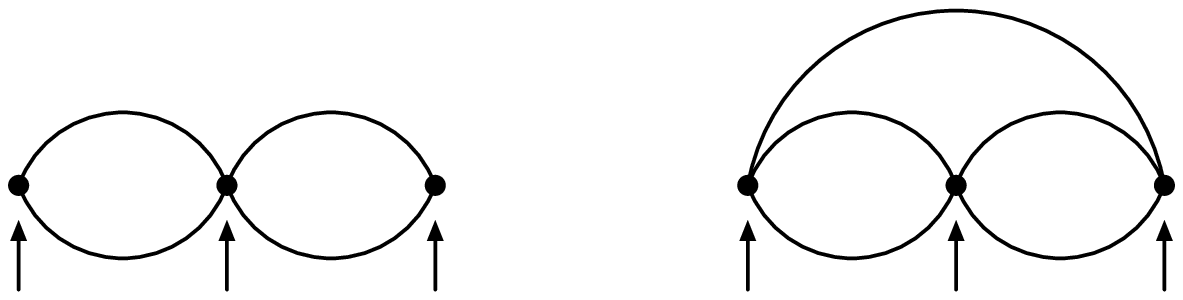}}
\point 30 0 {\mbox{destroy\ } B}
\point 90 0 {O^q_i, T^q_i}
\point 150 0 {\mbox{create\ } B}
\point 240 0 {\mbox{destroy\ } \Lambda_b}
\point 300 0 {O^q_1, T^q_1}
\point 360 0 {\mbox{create\ } \Lambda_b}
\end{picture}
\end{center}
\caption[]{Lattice quark propagator contractions needed to compute
spectator effects on $B$ and $\Lambda_b$ lifetimes.}
\label{fig:contractions}
\end{figure}
\begin{equation}
\begin{array}{rcrrcr}
B_1 &=& 1.06(8) & \qquad \epsilon_1 &=& -0.01(3)\\
B_2 &=& 1.01(6) & \qquad \epsilon_2 &=& -0.02(2)
\end{array}
\end{equation}
for the $B$ and
\begin{equation}
L_1 = \cases{-0.31(3) & A\\
             -0.22(4) & B\\}
\qquad
L_2 = \cases{0.23(2) & A\\
             0.17(2) & B\\}
\end{equation}
for the $\Lambda_b$. Combining these with the calculated coefficients
$a_i$ and $b_i$ leads to,
\begin{equation}
{\tau(B^-)\over \tau(B^0_d)} = 1.03(2)(3),\qquad
{\tau(\Lambda_b)\over \tau(B^0_d)} =
 \cases{0.91(1) & A \\ 0.93(1) & B\\}
\end{equation}
We see that the answer is close to unity as expected for the $B$ meson
ratio. For the $\Lambda_b$ to $B$ ratio, about $40\%$ of the deviation
from unity required to match experiment is reproduced. Given the
exploratory nature of the $\Lambda_b$ calculation, this indicates that
$1/m_b^3$ spectator effects \emph{could be} large enough to suppress
the $\Lambda_b$ lifetime compared to the $B$ meson lifetime and
further lattice studies are warranted. Moreover, next-to-leading order
calculations of the coefficient functions are in
progress~\cite{almb}. Experience with the $B_s$ lifetime difference,
where the next-to-leading effects are significant, provides further
motivation.

\section{Width Difference of $B_s$ Mesons}

The decay width difference of $B_s$ and $\bar B_s$ mesons arises from
the forward off-diagonal matrix element of a time ordered product of
weak effective Hamiltonians mediating $b$ quark decay. Applying the
heavy quark expansion leads to an expression of the
form~\cite{bbd,bbgln},
\begin{equation}
\fl \dgbs = {G_{\mathrm{F}}^2 m_b^2\over 12\pi M_{B_s}}\,
        |V_{cb}V_{cs}|^2
     \left( G(z)\langle Q_{L_{s}}(m_b)\rangle -
               G_S(z) \langle Q_S(m_b)\rangle +
               \hat\delta_{1/m}\sqrt{1-4z} \, \right)
\end{equation}
where $z=m_c^2/m_b^2$ and angle brackets denote matrix elements
between $B_s$ and $\bar B_s$, $\langle O\rangle = \langle\bar
B_s|O|B_s\rangle$. The four-quark $\Delta B = 2$ operators
$Q_{L_{s}}$ and $Q_S$
arise at leading order in the heavy quark expansion, while
$\hat\delta_{1/m}$ denotes $1/m_b$ corrections involving further
operator matrix elements. The coefficients $G$ and $G_S$ are known at
NLO in QCD~\cite{bbgln}. $Q_{L_{s}}$ is the same single operator that
contributes to the mass difference, $\Delta M_{B_s}$, while 
\begin{equation}
Q_S = (\bar b s)_{S-P}(\bar b s)_{S-P} .
\end{equation}
Notice that $\la Q_{L_{s}}\ra$ is actually the matrix element 
${\mathcal{M}}_{B_{s}}$ defined in Eq.~\eref{eq:bparamdef}. Using the
standard parameterisation,
\begin{equation}
\langle Q_S(\mu) \rangle =
     -{5\over3}\,\fbs^2 \MBs^2 {\MBs^2\over(m_b + m_s)^2}B_S(\mu) ,
\end{equation}
and Eq.~\eref{eq:bparamdef}, one can write,
\begin{eqnarray}
\label{eq:dgogfb}
\fl
\dgog = {G_{\mathrm{F}}^2 m_b^2\over 12\pi}\,
        |V_{cb}V_{cs}|^2 \, \tau_{B_s} \, \fbs^2 M_{B_s} \nonumber\\
      \times
        \left( {8\over3}\,G(z)B_{B_{s}} +
               {5\over3}\,G_S(z) {\MBs^2\over(m_b + m_s)^2} B_S +
               \delta_{1/m}\sqrt{1-4z} \, \right).
\end{eqnarray}
The coefficients $G_{(S)}$ and matrix element parameters $B_{B_{s}}$ 
and $B_{S}$
depend on a renormalisation scale $\mu$ such that the result is in
principle $\mu$ independent. Results quoted below have $\mu$ set to
the $b$-quark pole-mass, $m_b$, taking $m_b=4.6\gev$. Note that
different authors make different choices for the definitions of the
quark masses appearing in the defining equation for $B_S$. Here they
are the pole masses.

Input from lattice QCD is needed for the values of $\fbs$ and the
parameters $B_{B_s}$ and $B_S$. From the lattice viewpoint, it is
convenient to trade uncertainty in determining
$\fbs$~\cite{cthd,mcneile} for the appearance of extra CKM factors by
considering $\dgbs/\Delta\MBs$. The mass difference is proportional to
$\langle Q_{L_{s}}\rangle$, so that the ratio $\dgbs/\Delta\MBs$
depends on the quantity
\begin{equation}
\mathcal{R}(m_b) = {\langle Q_S\rangle\over\langle Q_{L_{s}}\rangle} =
 -{5\over8}\, {B_S(m_b)\over B_{B_{s}}(m_b)}\, {\MBs^2\over(m_b+m_s)^2}.
\end{equation}
Systematic errors from uncertainty in determining the lattice spacing
and from quenching effects should be reduced in this dimensionless
ratio of similar matrix elements.  In the absence of an experimental
measurement for $\Delta\MBs$, one further uses the ratio $\xi$, as
defined in Eq.~\eref{eq:deltamsovermd}, which is quite well determined by
lattice calculations (see \sref{sec:dmds}). In this way one arrives at
an expression~\cite{beci}
\begin{equation}
\label{eq:dgogratio}
\fl
\dgog = {4\pi\over3}\,{\MBs m_b^2\over M_{B_d} m_W^2}\,
        \left|{V_{cb}V_{cs}\over V_{tb}V_{td}}\right|^2 
        {\tau_{B_s} \Delta M_{B_d}\over \eta_B(m_b)S_0(x_t)}
        \xi^2 \left( G(z) - G_S(z) \mathcal{R}(m_b) +
                     \tilde\delta_{1/m} \right).
\end{equation}
The quantities $\eta(m_b)$ and $S_0(x_t)$, where $x_t = m_t^2/m_W^2$,
come from the expression for $\Delta\MBs$ and are known
factors~\cite{InamiLim,buras:1990fn}.  At leading order in $1/m_b$ the
non-perturbative contribution is isolated in $\mathcal{R}(m_b)$.  At
order $1/m_b$, one needs both the explicit $\tilde\delta_{1/m}$ piece
together with the implicit $m_b$ dependence from the matrix elements
of $Q_{L_{s},S}$ in the ratio $\mathcal{R}$.

Three groups have recent calculations giving values for $\mathcal
R$. They use different lattice formalisms, but give very consistent
results for this dimensionless ratio. Becirevic et al.~\cite{beci} use
heavy quarks at around the charm mass and extrapolate to the $b$. The
Hiroshima-KEK group~\cite{Hiroshima} use lattice NRQCD, simulating
directly at the $b$ quark mass. Finally, Gim\'enez and
Reyes~\cite{gra,grb} work in the lattice static quark theory, and
hence include $1/m_b$ corrections to the matrix elements as a
systematic error. All groups have results from quenched simulations,
but Gim\'enez and Reyes~\cite{grb} also have results with two flavours
of degenerate sea quarks.
\begin{equation}
\mathcal R (m_b) = \cases{%
  -0.93(3)({}^0_1)& extrap $c\to b$~\cite{beci}\\
  -0.91(5)(17)& NRQCD~\cite{Hiroshima}\\
  -0.95(7)(9)& static $b$~\cite{gra,grb}\\
  -0.97(5)(15)& static $b$, $n_f=2$~\cite{grb}\\}
\end{equation}
The variation of about $\pm3\%$ in the central values for $\mathcal R$
is smaller than the corresponding uncertainty of about $\pm10\%$ in
the values of $B_{B_s}$ and $B_S$.

Since the ratio $G(z)/G_S(z)$ is $3$--$4$\%~\cite{bbgln}, the
contribution from $\langle Q_S\rangle$ dominates that from $\langle
Q_{L_{s}}\rangle$ in $\dgbs/\Gamma_{B_s}$. However, the $1/m_b$
corrections, currently estimated using factorisation, cancel
significantly against the $\langle Q_S\rangle$ term~\cite{mbdurham},
so that two terms of order $0.1$--$0.15$ combine to give a prediction
$\dgbs/\Gamma_{B_s} \simeq 0.05$ from equation Eq.~\eref{eq:dgogratio}
with a large uncertainty. Moreover, using results for $B_{(B_s,S)}$ and
$\fbs$ in Eq.~\eref{eq:dgogfb} instead of Eq.~\eref{eq:dgogratio} gives
$\dgbs/\Gamma_{B_s}$ closer to $0.1$~\cite{mbdurham}: here there is
extra uncertainty from the lattice determination of $\fbs$, although
the $\delta_{1/m}$ uncertainty dominates. To make progress, better
input for the matrix elements in $\delta_{1/m}$ is vital, together
with better knowledge of $\fbs$: both may be addressed by future
lattice calculations. Note also that the `ratio' form of
Eq.~\eref{eq:dgogratio} assumes that $\Delta\MBs$ (and $\Delta M_{B_d}$)
are given by their standard model expressions. $\dgbs$ by itself is
dominated by tree level physics and so is expected to be less
sensitive to new physics than the mass differences. This would favour
working with the expression in Eq.~\eref{eq:dgogfb} once $\fbs$ is better
known.

\ack 

We thank Damir Becirevic, Laurent Lellouch and Chris Sachrajda warmly
for discussions and acknowledge PPARC grant
PPA/G/O/1998/00525 for support.  We are grateful to Laurent Lellouch
and Chris Sachrajda for reading the manuscript carefully.

\section*{References}

\addcontentsline{toc}{chapter}{Bibliography}
\bibliographystyle{prsty}
\bibliography{refs}

\end{document}